\documentclass[12pt,a4paper]{article}

\usepackage{amstex}
\usepackage{epsfig}
\usepackage{epic}

\newcommand{\bl}{{B-L}}

\newcommand{\GeV}{\,\mathrm{GeV}}
\newcommand{\MeV}{\,\mathrm{MeV}}

\newcommand{\la}[1]{\boldsymbol{\lambda}_{\sect{#1}}}

\newcommand{\g}{{\mathrm{G}}}
\newcommand{\p}{{\mathrm{P}}}
\newcommand{\s}{S}

\newcommand{\cp}{{\text{CP}}}
\newcommand{\gr}[1]{{\boldsymbol{#1}}} %% group representation

\newcommand{\ord}[1]{\mathcal{O}\left( #1 \right)}
\newcommand{\ordf}[2]{\mathcal{O}\fracwithdelims(){#1}{#2}}
\newcommand{\Vev}[1]{\left\langle #1\right\rangle}
\newcommand{\vev}[2][]{#1\langle #2 #1\rangle} %% \big\Big\bigg\Bigg  

\newcommand{\re}{\operatorname{Re}}

\newcommand{\interskip}{\medskip}
\newcommand{\sect}[1]{{\mathnormal{#1}}}
\newcommand{\repr}[1]{{#1}}

\newcommand{\nohyphens}%
        {\hyphenpenalty=10000\exhyphenpenalty=10000\relax}
\newcommand{\capdef}{}
\newcommand{\mycaption}[2][\capdef]{\renewcommand{\capdef}{#2}%
        \caption[#1]{{\small #2}}} 
\makeatletter
\renewcommand{\fnum@table}{\textbf{\tablename~\thetable}}
\renewcommand{\fnum@figure}{\textbf{\figurename~\thefigure}}
\makeatother

\newcounter{myenumi}

\renewcommand{\themyenumi}{\roman{myenumi}}
{\end{list}}

\newlength{\myem}
\newcounter{mysubequation}[equation]
\renewcommand{\themysubequation}{\alph{mysubequation}}
\newcommand{\mytag}{\stepcounter{mysubequation}%
        \tag{\theequation\makebox[\myem][c]{\themysubequation}}}
\newcommand{\restore}{\stepcounter{equation}}

\makeatletter
\renewcommand{\section}{\@startsection{section}{1}{0em}{2\baselineskip}%
{1\baselineskip}{\bf}}
\makeatother

\begin{document}

\vspace*{4\baselineskip}
\renewcommand{\thefootnote}{\fnsymbol{footnote}}
\noindent
January 1997 \hfill
IFUP--TH 5/97
\vspace*{5\baselineskip}\\
\noindent
\textbf{UNIFIED THEORIES OF FLAVOUR\\ WITH U(2) AS HORIZONTAL GROUP\footnote{Talk given at NATO Advanced Study Institute on Masses of Fundamental Particles,
Cargese, France, 5-17 Aug 1996.}}
\vspace*{3\baselineskip}\\
\noindent
\hspace*{1in} Andrea Romanino
\vspace*{1\baselineskip}\\
\hspace*{1in} Physics Department, University of Pisa\\
\hspace*{1in} and INFN, Sez.\ di Pisa,\\ 
\hspace*{1in} I-56126 Pisa, Italy
\vspace*{2\baselineskip}\\
\renewcommand{\abstractname}{ABSTRACT}
\small
\begin{abstract}\noindent
% \begin{minipage}{0.85\textwidth}
An unified flavour model based on U(2) as flavour group is
described. Besides to explain the main characteristics of the fermion
spectrum, the model is predictive and agrees quantitatively with
experimental data in flavour physics.
\end{abstract}
% \end{minipage}
\normalsize
\setcounter{footnote}{0}
\renewcommand{\thefootnote}{\arabic{footnote}}
\section*{THE U(2) SYMMETRY}

The aim of this exposition is to describe a theory of flavour based on
SU(5) or SO(10) as gauge group and U(2) as flavour group. Such an
unified theory has been developed by R.\ Barbieri, L.~J.\ Hall, S.\
Raby and myself~\cite{BHRR} on the basis of previous works by R.\
Barbieri, G.\ Dvali and L.J.\ Hall~\cite{barbieri:96b} and R.\
Barbieri and L.~J.\ Hall~\cite{BH}.

While the structure of the gauge sector of the Standard Model (SM) is
well understood in terms of the unification hypothesis, a coherent and
complete quantitative explanation of the flavour sector, in particular
of fermion masses and mixings is still
missing. Ref.~\cite{anderson:94a} is an attempt in this
direction. However, besides the bottom quark and $\tau$ lepton mass
unification~\cite{chanowitz:77a,ibanez:83a}, several relations have
been noticed in the past, sometimes justified on a theoretical basis,
like, e.g., $|V_{us}|\simeq(m_d/m_s)^{1/2}$~\cite{gatto:68a},
$m_{\mu}\simeq 3 m_s$ and $m_d\simeq 3 m_e$~\cite{georgi:79a}, or
$|V_{ub}/V_{cb}|\simeq(m_u/m_c)^{1/2}$~\cite{harvey:80a}, involving
the masses and the CKM matrix elements renormalized at the unification
scale.  In a supersymmetric theory, the flavour problem extends to scalar
masses and scalar-fermion mixings, that have to be consistent with
experimental limits on Flavour Changing Neutral Current (FCNC)
phenomena~\cite{gabbiani:96a}.

In order to explain the ``horizontal'' relations among masses and
mixings, table~\ref{tab:relazioni}, in terms of coupling constants of
order one, we introduce as usual an horizontal symmetry acting in the
same way on the family indexes of different SM
representations\footnote{As it is if the symmetry group commutes with a
fully unified group.}, and we suppose that this symmetry is
spontaneously broken by the SM-invariant vacuum expectation values of
``flavon'' fields.
\begin{table}
\mycaption{Fermion masses and mixing relations at the unification scale.}
\label{tab:relazioni}
\[
\begin{array}{|c|c|c|c|}
% 	  
% 	& \mathbf{3}
% 	& \mathbf{2/3}% \vspace*{2\baselineskip}

% 	& \mathbf{12/33} \\
\hline
	  \sect{U}/\sect{D},\sect{E}
	& m_t \gg m_b, m_\tau
	& \raisebox{0ex}[3.5ex][2.8ex]{$\displaystyle\frac{m_c}{m_t}
	  \ll \frac{m_s}{m_b}, 
	  \frac{m_\mu}{m_\tau}$}
	& \raisebox{0ex}[3.5ex][2.8ex]{$\displaystyle\frac{m_u
	  m_c}{m_t^2} \ll \frac{m_d 
	  m_s}{m_b^2}, \frac{m_e m_\mu}{m^2_\tau}$} \\
\hline
	  \sect{D}/\sect{E}
	& m_b \approx m_\tau
	& \raisebox{0ex}[3.5ex][2.8ex]{$\displaystyle 3
	  \frac{m_s}{m_b} \approx \frac{m_\mu}{m_\tau}$} 
	& \raisebox{0ex}[3.5ex][2.8ex]{$\displaystyle \frac{m_d
	  m_s}{m^2_b} \approx \frac{m_e m_\mu}{m^2_\tau}$} \\
\hline
\hline
	  V_{\textrm{CKM}}
	& \raisebox{0ex}[3.65ex][2.8ex]{$\displaystyle |V_{us}| \sim
	  \sqrt{\frac{m_d}{m_s}}$}
	& \raisebox{0ex}[3.65ex][2.8ex]{$\displaystyle |V_{cb}| \sim
	  \frac{m_s}{m_b}$}
	& \raisebox{0ex}[3.65ex][2.8ex]{$\displaystyle 
	  \fracwithdelims||{V_{ub}}{V_{cb}} \sim
	  \sqrt{\frac{m_u}{m_c}}$} \\
\hline
\end{array}
\]
\end{table}

Instead of considering the possible horizontal symmetries involving
all the three families, we limit ourselves to symmetries acting on the
two lightest families. Moreover we require that these families are
massless in the fermion sector and exactly degenerate in the scalar
sector in the limit of unbroken symmetry, as suggested by the
smallness of light fermion Yukawa couplings and by the upper limits on
FCNC phenomena. The only suitable continuous non abelian unitary group
is then U(2). We prefer to consider a non abelian group because the
representation structure, and so the theory, is more constrained than
in the abelian case, where there is a large freedom in choosing the
horizontal quantum numbers.  On the contrary, in the U(2) case, the
transformation properties of the flavons can be guessed from the
requirement of having mixing among the three fermion families.

We do not want direct couplings of the flavon fields $\phi$ to the
fermion fields $\psi$ through renormalizable operators. Rather, we
suppose that the flavons appear only in non renormalizable effective
operators generated below a ``flavour scale'' $M=\ord{M_\g\div M_\p}$
at which the physics that mediate the U(2) breaking is integrated
out. An example of such a physics is given by the ``Froggatt-Nielsen
mechanism'', as explained by G. Ross in his
lectures~\cite{froggatt:79a}. The ratios $\vev{\phi}/M$ provide the
small parameters needed to describe the fermion mass and mixing
structure without using small couplings.

Let us choose the transformation properties of the two light families,
$\psi_a$, $a=1,2$, under U(2):
\[
U\in \mathrm{U(2)}:\; \psi_a\rightarrow U_a^b \psi_b
\]
(the third family, as the Higgses, is supposed to be U(2)-invariant).
Then, in order to generate mixing between the third and the light
families, it is necessary to have a flavon field $\phi^a$ that transforms
in the conjugate way relative to $\psi_a$. This flavon gives rise to
12/3 mixing through Yukawa interactions like
$(\vev{\phi^a}/M)\psi_a\psi_3 h$. But $\phi^a$ only breaks U(2) to
U(1) (at a scale $V$ with $V/M=\ord{|V_{cb}|_\g}=\ord{0.02}$) and the
U(1) residual symmetry prevents the lightest generation to get mass
and to mix with the heavier ones. In order to break U(1) and to
generate the operator $\psi_a\psi_b h$, it is necessary to have a flavon
field $\phi^{ab}=A^{ab}+S^{ab}$ with transformation properties under
U(2) as indicated by the position of the indexes, where $A^{ab}$ and
$S^{ab}$ are their antisymmetric and symmetric irreducible
components. Since $\vev{A^{ab}}$ breaks U(1) and gives mass to the
lightest family, the corresponding scale of breaking $v$ has to be lower
than $V$. Moreover, in order to generate the $m_s/m_b$ ratio at the
same scale of $|V_{cb}|$, $S^{ab}$ should break U(2) at the $V$
scale. If we finally suppose that all breakings of U(2) at the $V$
scale leave unbroken the same U(1), we are left with the following
breaking pattern (in a suitable basis)
\begin{gather}
\label{rottura}
\text{U(2)} @>{\Vev{\phi^a},\Vev{S^{ab}}}>> \text{U(1)}
@>{\Vev{A^{ab}}}>> \{\mathbf{1}\} \mytag \\[0.2\baselineskip]
\vev{\phi^a} = \begin{pmatrix} 0 \\ V \end{pmatrix}, \enspace 
\vev[\big]{A^{ab}} = \begin{pmatrix} 0 & v \\ -v & 0 \end{pmatrix}, \enspace 
\vev[\big]{S^{ab}} = \begin{pmatrix} 0 & 0 \\ 0 & V' \end{pmatrix},
\mytag \\ 
V,v>0, \qquad v \ll V \approx |V'| \ll M. \mytag
\end{gather}
\restore
Also U(2) invariant fields can develop an expectation value and play a
role. We denote them with $\Sigma$.

As a consequence of the previous breaking pattern, the Yukawa matrices
$\la{}$ and the scalar masses $\mathbf{m}^2$ are in the
form
\begin{equation}
\label{matrices}
\la{} = \begin{pmatrix}
0 & \epsilon' & 0 \\
-\epsilon' & \epsilon & x\epsilon \\
0 & y\epsilon & 1 \end{pmatrix} \qquad
\mathbf{m}^2 = \begin{pmatrix}
m^2_1 & 0 & 0 \\
0 & m^2_1(1+z\epsilon^2) & \epsilon {m^2_2}^* \\
0 & \epsilon m^2_2 & m^2_3 \end{pmatrix}
\end{equation}
where $\epsilon \equiv V/M$, $\epsilon'\equiv v/M$,
$x,y,z=\ord{1}$ and $m^2_1,m^2_2,m^2_3=\ord{m^2_\s}$. This leads to a CKM matrix in the form~\cite{hall:93a}
\begin{equation}
\label{VS}
V_{\text{CKM}} = \begin{pmatrix}
c^{\sect{D}}_{12} & s^{\sect{D}}_{12} -s^{\sect{U}}_{12} e^{i\phi} &
s^{\sect{U}}_{12}s\\
s^{\sect{U}}_{12} -s^{\sect{D}}_{12} e^{i\phi} & c^{\sect{D}}_{12} e^{i\phi}
& -c^{\sect{U}}_{12}s \\
-s^{\sect{D}}_{12} s  & c^{\sect{D}}_{12}s &
 e^{-i\phi}
\end{pmatrix}
\end{equation}
in terms of 
\begin{equation}
\label{angdef}
t^D_{12} = \frac{s^D_{12}}{c^D_{12}} = \sqrt{\frac{m_d}{m_s}} \qquad
t^U_{12} = \frac{s^U_{12}}{c^U_{12}} = \sqrt{\frac{m_u}{m_c}} 
\end{equation}
and of the parameters $s>0$ and $\phi\in(0,\pi)$.

The previous form of the CKM matrix is in agreement with the
experimental values of $|V_{us}|$ and $|V_{ub}/V_{cb}|$. Moreover,
due to the $\epsilon$ term in the $\lambda_{22}$ entry, in the down
quark sector we have
\begin{equation}
\label{ok}
\frac{m_s}{m_b} = \ord{\epsilon} = \ord{|V_{cb}|} \quad\text{and}\quad
\frac{\tilde{m}^2_d - \tilde{m}^2_s}{\tilde{m}^2_d}\leq
\ord{\epsilon^2} = \ordf{m_s^2}{m_b^2},
\end{equation}
in agreement with the experimental values of $m_s/m_b$ and $|V_{cb}|$
and with the limits on $(\tilde{m}^2_d - \tilde{m}^2_s)/\tilde{m}^2_d$
given by the value of $\epsilon_K$, the CP violation parameter in the
$K^0$-$\bar{K}^0$ system~\cite{gabbiani:96a}.  In models in which the
22 entry is vanishing or of order $\epsilon^2$ it is on the contrary
\begin{equation}
\label{no}
\frac{m_s}{m_b} = \ord{\epsilon^2} = \ord{|V_{cb}|^2} \qquad
\frac{\tilde{m}^2_d - \tilde{m}^2_s}{\tilde{m}^2_d}\leq
\ord{\epsilon^2} = \ordf{m_s}{m_b}.
\end{equation}

The CKM matrix in~\eqref{VS} has been obtained by diagonalizing a
Yukawa matrix in the form of eq.~\eqref{matrices} in each sector of
given charge $U$, $D$, $E$ without comparing the Yukawa matrices of
different sectors. By doing that, we immediately see that, to
reproduce the suppression of the mass ratios in the $U$-sector
relative to the $D$- and $E$-sector ones (``$U/D,E$'' line in
table~\ref{tab:relazioni}), the 12, 21 and 22 entries of the
$U$-sector Yukawa matrix, $\la{U}$, have to be suppressed relative to
the corresponding entries in $\la{D}$ and $\la{E}$. This suppression
can be explained in terms of couplings of order one only in presence
of vertical relations among the coupling constants as those ones
provided by an unified gauge group.

\section*{UNIFICATION}

Let us then consider the U(2) theory in the case of the minimal unified
group SU(5), where each family is represented by a tenplet $T$ and an
anti-fiveplet $\bar{F}$. The flavons must transform as $\gr{1}$,
$\gr{24}$ or $\gr{75}$ under SU(5), otherwise they cannot have a
SM-invariant expectation value or they cannot contribute to the
Yukawa matrices linearly. The only renormalizable interactions
are then
\begin{equation}
  \label{int0}
  T_3 T_3 H, \qquad \bar{F}_3 T_3 H,
\end{equation}
while the possible non renormalizable interactions at the linear order
in the flavons are 
\begin{gather}
  \label{int1}
\bar{F}_3 T_a\bar{H}\frac{\phi^a}{M}, \qquad
\bar{F}_a T_3\bar{H}\frac{\phi^a}{M}, \qquad
T_3 T_a H\frac{\phi^a}{M}, \mytag \\
T_a T_b H\frac{S^{ab}}{M}, \qquad
T_a T_b H\frac{A^{ab}}{M}, \mytag \\
\bar{F}_a T_b H\frac{S^{ab}}{M}, \qquad
\bar{F}_a T_b H\frac{A^{ab}}{M}. \mytag
\end{gather}
\restore
The operators $T_a T_b H \vev{S^{ab}}/M$ and $T_a T_b H
\vev{A^{ab}}/M$ are responsible for the contributions to $m_c/m_t$ and
$(m_u m_c/m_t^2)^{1/2}$ at linear order in $\epsilon$, $\epsilon'$
respectively. The $m_c/m_t$ and $(m_u m_c/m_t^2)^{1/2}$ suppressions
can then be explained if this two operators are vanishing and these
ratios are generated at second order in the expectation
values. Can this happen in a natural way? Let us consider first the
operator $T_a T_b H \vev{S^{ab}}/M$. Since only the symmetric part of
the product $T_aT_b$ participates in this interaction, if $S^{ab}$
transforms as $\gr{75}$ under SU(5) the operator $T_a T_b H S^{ab}/M$
vanishes because it cannot be constructed in a SU(5) invariant way.
If $S^{ab}$ is not a $\gr{75}$ of SU(5), the only case in which $T_a
T_b H \vev{S^{ab}}/M$ vanishes is when $S^{ab}$ has two components
transforming as $\gr{1}$ and $\gr{24}$ respectively, but only for a
particular value of the ratio of their expectation values. This
possibility has not a natural interpretation in the SU(5) context but
has to be taken in consideration when additional vertical structure is
present, as it is the case in SO(10).

Analogously, the $T_a T_b H \vev{A^{ab}}/M$ operator vanishes only
if $A^{ab}$ is a SU(5) singlet or if $A^{ab}$ has two components
transforming as $\gr{24}$ and $\gr{75}$, for a particular value of the
ratio of their expectation values.

Thus the SU(5)$\times$U(2) theory allows a natural explanation of the
suppression of the $U$-sector mass ratios that requires $S^{ab}\sim
\gr{75}$ and $A^{ab}\sim \gr{1}$. It is very interesting
that in this case the $D/E$ relations in table~\ref{tab:relazioni} are
\emph{automatically} predicted as a consequence of 
$\lambda^D_{33}=\lambda^E_{33}$,
$\lambda^D_{22}=-\frac{1}{3}\lambda^E_{22}$,
$\lambda^D_{12}=\lambda^E_{12}$ that follow from the SM decomposition
of the operators  $\bar{F}_a T_b \bar{H}\vev{S^{ab}}/M$ and $\bar{F}_a
T_b \bar{H}\vev{S^{ab}}/M$.

Let us consider now full unification of every single families through
SO(10). In this case, the possible representations of the flavons are
$\gr{1}$, $\gr{45}$, $\gr{54}$ and $\gr{210}$. It is possible to
implement the mechanism described above by using representations for
$A^{ab}$ and $S^{ab}$ having components transforming as $\gr{1}$ and
$\gr{75}$ respectively under SU(5). In this case, $S^{ab}$ must
transform as $\gr{210}$ under SO(10).

On the other hand, it is possible to explain the suppression of the
$U$-sector mass ratios using only singlets and adjoints of
SO(10). This is possible if $S^{ab}$ is a SO(10) adjoint and if the
particular value of the ratio of its components transforming as
$\gr{1}$ and $\gr{24}$ under SU(5) is the one characteristic of the
$B-L$ generator. Moreover the expectation value of $A^{ab}$ must be
SU(5) invariant.

To generate higher order contributions to the fermion masses we use
the U(2)-invariant SO(10)-adjoint field $\Sigma$.  If we decompose its
breaking $\vev{\Sigma}$ along the generators $X$ and $Y$,
$\vev{\Sigma}=\Sigma_X+\Sigma_Y$, ($X$ and $Y$ generate the
SM-invariant generator subspace and are given in
table~\ref{tab:generatori}) we see that the $D$- and $E$-sector light
masses require a ratio $\vev{\Sigma_X}/M$ not far from 1, while
$\vev{\Sigma_Y}/M=\ord{0.02}$ is responsible for the suppression of
the $U$-sector mass ratios.  Also in this case the $D/E$ relations in
table~\ref{tab:relazioni} are automatically predicted, but
$\lambda^D_{22}=-\lambda^E_{22}/3$ becomes
$\lambda^D_{22}=\lambda^E_{22}/3$. The relevant operators are then
\begin{gather}
\label{int10}
\psi_3 f_1 \left( {\Sigma_X \over M} \right) H \psi_3 \mytag\\
{1 \over M} \psi_3 \vev{\phi^a} f_2 \left( {\Sigma_X \over M} \right)
H \psi_a \mytag\\ {1 \over M} \psi_a \left(\vev[\big]{S^{ab}} f_3
\left( {\Sigma_X \over M} \right) + \vev[\big]{A^{ab}} f_4 \left(
{\Sigma_X \over M} \right) \right)H \psi_b
\mytag\\ 
\begin{gathered}
{1 \over M^2} \psi_a \left(\vev[\big]{S^{ab}} \Sigma_Y f_5 \left(
      {\Sigma_X \over M} \right) + \vev[\big]{A^{ab}} \Sigma_Y f_6
      \left( {\Sigma_X \over M} \right)  \right)H \psi_b \\
+{1 \over M^2} \psi_a \vev{\phi^a}\vev{\phi^b} f_7 \left(
      {\Sigma_X \over M} \right)H \psi_b, 
\end{gathered}
\mytag
\end{gather}
\restore
where the functions $f$ take into account all orders in
$\Sigma_X/M$ and we only consider the leading orders in
$\Sigma_Y/M$.
\begin{table}
    \mycaption{Values of some SM invariant generators on SM
      representations. $X$ is SU(5) invariant, $Y$ is the hypercharge,
      $B-L$ the barion minus lepton number and $T_{3R}$ the third
      component of the right weak isospin.}
      \label{tab:generatori}
\centering
\[
\begin{array}{c|cccccc}
    
  & \repr{Q}
  & \repr{u}
  & \repr{d}
  & \repr{L}
  & \repr{\nu}
  & \repr{e} \\
\hline
    X
  & 1
  & 1
  & -3
  & -3
  & 5
  & 1 \\
    Y
  & 1/6
  & -2/3
  & 1/3
  & -1/2
  & 0
  & 1 \\
    \bl
  & 1/3
  & -1/3
  & -1/3
  & -1
  & 1
  & 1 \\
    T_{3R}
  & 0
  & 1/2
  & -1/2
  & 0
  & 1/2
  & -1/2
\end{array}
\]
\end{table}

\section*{PREDICTIONS OF UNIFIED MODELS}

In the unified models described above, the Yukawa matrices are of the
form
% \begin{gather}
%   \label{Yukawaunif}
% \la{U} = 
% \begin{pmatrix}
%   0 & \epsilon'\rho & 0 \\
% -\epsilon'\rho & \epsilon\rho' & y_{\sect{U}}\epsilon \\
% 0 & x_{\sect{U}}\epsilon & 1
% \end{pmatrix} \lambda
% \mytag \\[0.3\baselineskip]
% \la{D} = 
% \begin{pmatrix}
%   0 & \epsilon' & 0 \\
% -\epsilon' & \epsilon & y_{\sect{D}}\epsilon \\
% 0 & x_{\sect{D}}\epsilon & 1
% \end{pmatrix} \lambda\zeta
% \mytag \\[0.3\baselineskip]
% \la{E}^T = 
% \begin{pmatrix}
%   0 & \epsilon' & 0 \\
% -\epsilon' & \pm 3\epsilon & y_{\sect{E}}\epsilon \\
% 0 & x_{\sect{E}}\epsilon & 1
% \end{pmatrix} \lambda\zeta,
% \mytag 
% \end{gather}
% \restore
\begin{equation}
\la{U} = 
\begin{pmatrix}
  0 & \epsilon'\rho & 0 \\
-\epsilon'\rho & \epsilon\rho' & y_{\sect{U}}\epsilon \\
0 & x_{\sect{U}}\epsilon & 1
\end{pmatrix} \lambda, \qquad
\la{(D,E)} = 
\begin{pmatrix}
  0 & \epsilon' & 0 \\
-\epsilon' & (1,\pm 3)\epsilon & y_{\sect{(D,E)}}\epsilon \\
0 & x_{\sect{(D,E)}}\epsilon & 1
\end{pmatrix} \lambda\zeta,
\label{Yukawaunif}
\end{equation}
where $x_{U,D,E}$, $y_{U,D,E}$, $\lambda$, $\rho'/\rho$ are of order
one, $\epsilon,\rho=\ord{0.02}$, $\epsilon'=\ord{0.004}$ and the
factor $\zeta=\ord{0.02}$ corresponds to the possibility that the two
light Higgs doublets appear in the unified multiplet in the
renormalizable Yukawa interaction with different weights. In this way
we do not solve the $m_b/m_t\ll 1$ problem but we reexpress it in term
of $\zeta\ll 1$.

From eq.~\eqref{Yukawaunif}, both qualitative and quantitative
predictions follow. If we neglect the order one parameters, we can
have order of magnitude expressions for the 13 Yukawa observables at
the unification scale in terms of 4 small parameters, $\epsilon$,
$\rho$, $\epsilon'$ and $\zeta$. This gives rise to 9 qualitative or
order of magnitude predictions in the Yukawa sector\footnote{The
$\ord{1}$ parameter neglected in $J_\cp$, $\sin\phi$, could in
principle be vanishing, making the prediction for the $J_\cp$ order of
magnitude wrong. Actually, from the $|V_{us}|$ value it turns out that
$\sin\phi=\ord{1}$, so that also the $J_\cp$ relation is correct.}:
\begin{gather}
  \label{preunif}
m_t \sim v = 174\GeV \mytag \\
m_b \sim m_\tau \quad
3\frac{m_s}{m_b} \sim \frac{m_\mu}{m_\tau} \quad
\frac{m_d m_s}{m_b^2} \sim \frac{m_e m_\mu}{m_\tau^2} \mytag \\
\frac{m_s}{m_b} \sim \fracwithdelims(){m_c}{m_t}^{1/2} \mytag \\
\fracwithdelims||{V_{ub}}{V_{cb}} \sim
\sqrt{\frac{m_u}{m_c}} \quad 
\fracwithdelims||{V_{td}}{V_{ts}} \sim
\sqrt{\frac{m_d}{m_s}}  \mytag \\
|V_{cb}| \sim \frac{m_s}{m_b} \qquad J_\cp \sim
\left(\frac{m_d}{m_s}\frac{m_u}{m_c}\right)^{1/2} |V_{cb}|^2, \mytag
\end{gather}
\restore
besides $|V_{us}|=\ord{\sqrt{m_d/m_s}}$.

Let us come now to the quantitative predictions. Taking into account
order one factors, from the diagonalization of the Yukawa matrices in
eqs.~\eqref{Yukawaunif} we obtain the 5 following relations:
\begin{gather}
  \label{preapp}
\frac{m_\mu}{m_\tau}\left(1-\frac{m_e}{m_\mu}\right) \simeq
3 \frac{m_s}{m_b}\left(1-\frac{m_d}{m_s}\right) \mytag \\  
m_b \simeq m_\tau \qquad \frac{m_e m_\mu}{m_\tau^2} \simeq 
\frac{m_d m_s}{m_b^2}  \mytag \\  
\fracwithdelims||{V_{ub}}{V_{cb}} \simeq 
\sqrt{\frac{m_u}{m_c}} \qquad 
\fracwithdelims||{V_{td}}{V_{ts}} \simeq
\sqrt{\frac{m_d}{m_s}}. \mytag
\end{gather}
\restore
Eq.~(\ref{preapp}a) has $\ord{\epsilon}$ corrections from the
diagonalization, while eqs.~(\ref{preapp}b,c) have only $\ord{\epsilon^2}$
corrections. All of them, however, have corrections due to possible
higher order contributions in $\Sigma_Y/M$ or to weak scale
radiative corrections, both of order $\ord{\epsilon}$.

The quantitative predictions in eqs.~\eqref{preapp} all agree with the
experimental values of the quantities involved. On the other
hand these errors are not so small. Nevertheless, as H. Leutwyler
explained by his lectures~\cite{leutwyler:95a}, there is a combination
of light quark masses that is known with good precision:
\begin{equation}
  \label{Q}
Q = \frac{m_s/m_d}{\sqrt{1-(m_u/m_d)^2}} = 22.7\pm0.08. 
\end{equation}
Taking into account $\ord{\epsilon}$ corrections to~(\ref{preapp}a) we
get our prediction
\begin{equation}
  \label{Qpre}
  Q \simeq \frac{25}{\sqrt{1-(m_u/m_d)^2}}\cdot (1-2\Delta),
\end{equation}
with 
\begin{equation}
  \label{s/dcorr}
  \Delta = \re\left[\left(x_{\sect{D}}y_{\sect{D}}\mp
  \frac{1}{3}x_{\sect{E}}y_{\sect{E}}\right)\epsilon\right].
\end{equation}
The $Q$ value, together with $|V_{cb}|=|(x_U-x_D)\epsilon|$, is then a
constraint on the $\ord{\epsilon}$ corrections, that depend on the 23
and 32 entries of the Yukawa matrices and, in turn, on the direction
of the $\phi^a$ expectation value in the generator space\footnote{A
$\phi$ singlet under SO(10) is excluded.}.

It is also possible to test quantitatively the class of considered
models by doing a fit of the CKM matrix. Eqs.~\eqref{VS}
and~\eqref{angdef}, that are pure U(2) consequences, allows a
parameterization of $V_{\text{CKM}}$ in terms of $s$ and $\phi$,
besides $m_u/m_c$ and $m_d/m_s$ that, fixing $Q$, $m_t$, $m_b$, $m_c$,
can be expressed in terms of $m_s$, $\alpha_s(M_Z)$, $m_u/m_d$. We can
then perform a fit of the U(2) model expressing the measured
$|V_{us}|$, $|V_{cb}|$, $|V_{ub}/V_{cb}|$, $\alpha_s(M_Z)$, $m_s$ in
terms of $m_u/m_d$, $s$, $\phi$, $\alpha_s(M_Z)$, $m_s$. In the
unified case, $m_s$ can be expressed in terms of lepton masses using
the relation $m_e m_\mu = m_d m_s$ valid at the unification
scale. Finally, if we assume that $\epsilon_K$ and the
$B_d$-$\bar{B}_d$ mass difference are accounted for by the usual SM
diagrams, we can further constrain the fit and extend it to the SM
case, in which only the parameterization in~\eqref{VS} is used.

\begin{figure}
\centering
\epsfig{file=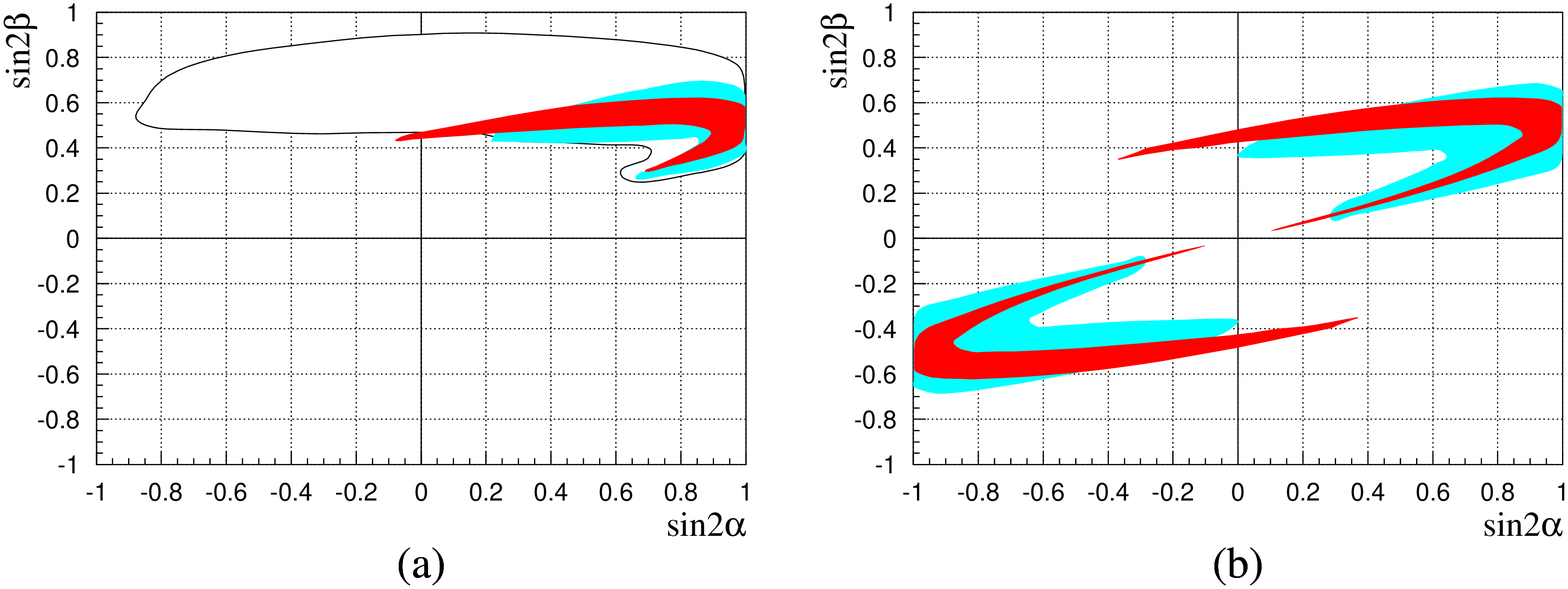,width=\textwidth}
\mycaption{Regions in the $\sin 2\alpha$--$\sin 2\beta$ plane, as
determined at the 90\% confidence level by the fit in the SM (white area in
  (a)), in the ``pure'' U(2) model (grey area) and in the unified U(2)
  model (dark area), in the ``constrained'', (a), and  ``not
  constrained'', (b), case.}
\label{fig:sinsth}
\end{figure}

The results of the fit in the unified U(2) models are summarized in
table~\ref{tab:bet}. Also, in figure~\ref{fig:sinsth} the predictions for
the angles $\alpha$, $\beta$ of the
unitarity triangle, 
% \begin{align}
% \label{unit}
% \alpha & =
% \arg\left(-\frac{V_{tb}^*V_{td}^{}}{V_{ub}^*V_{ud}^{}}\right) 
% \mytag \\
% \beta  & =
% \arg\left(-\frac{V_{cb}^*V_{cd}^{}}{V_{tb}^*V_{td}^{}}\right) 
% \mytag \\
% \gamma & =
% \arg\left(-\frac{V_{ub}^*V_{ud}^{}}{V_{cb}^*V_{cd}^{}}\right) 
% \mytag
% \end{align}
% \restore
\begin{equation}
\label{unit}
\alpha =
\arg\left(-\frac{V_{tb}^*V_{td}^{}}{V_{ub}^*V_{ud}^{}}\right) \qquad
\beta  =
\arg\left(-\frac{V_{cb}^*V_{cd}^{}}{V_{tb}^*V_{td}^{}}\right) \qquad
\gamma =
\arg\left(-\frac{V_{ub}^*V_{ud}^{}}{V_{cb}^*V_{cd}^{}}\right) 
\end{equation}
are shown. In terms of the parameterization of the CKM matrix in~\eqref{VS},
these angles turn out to be 
\begin{align}
  \label{unitpar}
\alpha &= \phi \\
\beta  &= \arg\left(1-\frac{s^{\sect{U}}_{12}}{s^{\sect{D}}_{12}}
  e^{-i\phi}\right) \\
\gamma &= \pi-\alpha-\beta.
\end{align}
\begin{table}
\renewcommand{\arraystretch}{1.3}
\mycaption{Results of the fit in the unified {\rm U(2)} theories with
            (``constrained'') or without (``unconstrained'') inclusion
of $\epsilon_K$ and $\Delta m_{B_d}$ in the inputs.}
\label{tab:bet}
\[
\begin{array}{||c|c|c|c||}
\hline
            &\mbox{inputs}
            &\mbox{constrained}
            &\mbox{unconstrained} \\
\hline
m_s/\MeV     
            & 175 \pm 55      
            & 153^{+35}_{-22}  
            & 153\pm 35 \\
|V_{cb}|    
            & 0.038 \pm 0.004 
            & 0.039^{+0.0025}_{-0.0015} 
            & 0.038 \pm 0.004 \\
|V_{ub}/V_{cb}| 
            & 0.08 \pm 0.02 
            & 0.075\pm 0.013
            & 0.075\pm 0.016 \\
\epsilon_K\cdot 10^3
            & 2.26
            & 2.26
            & \pm (1.7^{+1.3}_{-0.1}) \\
B_K               
            & 0.8 \pm 0.2 
            & 0.86\pm 0.16 
            & 0.8 \\

\Delta m_{B_d}/\mbox{ps}^{-1}
            & 0.464
            & 0.464
            & 0.37^{+0.14}_{-0.05} \\
\sqrt{B}f_B/\MeV  
            & 200 \pm 40  
            & 178\pm 18
            & 200 \\
\hline
\alpha_s(M_Z)          
            & 0.117 \pm 0.006 
            & 0.118\pm 0.005 
            & 0.118\pm 0.005 \\
\hline
\end{array}
\]
\end{table}
\interskip

In conclusion, I briefly summarize the characteristics of the unified
U(2) models that we have considered.
\begin{itemize}
\item The models are simple and motivated.
\item The order of magnitude of 13 fermion masses and mixings are
qualitatively understood in terms of 4 small parameters: the ratios of
the two U(2) breaking scales and the flavour scale $M$, the ratio of
the SU(5) breaking scale and $M$ and the ratio of the coefficients of
the light Higgs doublets in the unified Higgs multiplet. The scales
are then as shown in figure~\ref{fig:scales}.
\begin{figure}
\footnotesize
\centering
\setlength{\unitlength}{0.45\textwidth}
\begin{picture}(2.05,0.825)(-1.05,-0.42)
\put(-1.05,-0.41){\framebox(1.0,0.825){}}
\put(-1.05,-0.41){\framebox(2.05,0.825){}}
\thicklines

\put(-0.90,0){\line(1,0){0.35}}
\put(-0.90,0.3){\line(1,0){0.35}}
\put(-0.90,-0.2){\line(1,0){0.35}}
\put(-1,0.3){$M$}
\put(-1,0.0){$M_G$}
\put(-0.51,0){$\vev{S}\simeq\vev{\phi}\simeq\vev{\Sigma_Y}$}
\put(-0.51,-0.20){$\vev{A}$}
% \put(-0.51,0.22){\makebox(0.35,0.08)[l]{[${\rm SO}_{10}\times{\rm U}_2$]}}
\put(-0.90,0.22){\makebox(0.35,0.08){${\rm SU}_5\times{\rm U}_2$}}
\put(-0.90,-0.08){\makebox(0.35,0.08){${\rm SU}_{3,2,1}\times{\rm U}_1$}}
\put(-0.90,-0.28){\makebox(0.35,0.08){${\rm SU}_{3,2,1}$}}
\put(-1.05,-0.4){\makebox(1,0.08){(a)}}

\put(0.15,0){\line(1,0){0.35}}
\put(0.15,0.2){\line(1,0){0.35}}
\put(0.15,0.3){\line(1,0){0.35}}
\put(0.15,-0.2){\line(1,0){0.35}}
\put(0.05,0.3){$M$}
\put(0.05,0.2){$M$}
\put(0.05,0.0){$M_G$}
\put(0.0,0.2){\makebox(0.05,0.1)[bl]{$\simeq$}}
\put(0.54,0){$\vev{S}\simeq\vev{\phi}\simeq\vev{\Sigma_Y}$}
\put(0.54,-0.20){$\vev{A}$}
\put(0.54,0.20){$\vev{\Sigma_X}$}
\put(0.15,0.22){\makebox(0.35,0.08){${\rm SO}_{10}\times{\rm U}_2$}}
\put(0.15,0.12){\makebox(0.35,0.08){${\rm SU}_5\times{\rm U}_2$}}
\put(0.15,-0.08){\makebox(0.35,0.08){${\rm SU}_{3,2,1}\times{\rm U}_1$}}
\put(0.15,-0.28){\makebox(0.35,0.08){${\rm SU}_{3,2,1}$}}
\put(-0.05,-0.4){\makebox(1.05,0.08){(b)}}
\normalsize

\end{picture}
\mycaption{Scales of symmetry breaking vevs appropriate to the SU(5)
(a) and SO(10) (b) cases described in the text.}
\label{fig:scales}
\end{figure}
\item A quantitative analysis of experimental data is possible in the
general model through a successful fit that predicts a strong
correlation among the $\alpha$ and $\beta$ angles of the unitarity
triangle. Moreover, 5 precise relations among mass and mixings are
predicted.
\item The suppression of $U$-sector mass ratios is explained in a
natural way and automatically leads to the $D/E$ relations in
table~\ref{tab:relazioni}. 
\item The scalar masses and mixings fulfill the requirements following
from the experimental limits on FCNC phenomena. Moreover, they
are constrained by U(2) symmetry and unification. This allows an
analysis of contributions from supersymmetric particles to several
quantities in flavour physics~\cite{progress}.
\end{itemize}

\renewcommand{\refname}{REFERENCES}

% \bibliographystyle{phaip}
% \bibliography{abbrev,biblio,tesi,con1,art4}

\end{document}